\documentstyle[graphicx,aps,prl,twocolumn,amssymb]{revtex}

\newcommand{\mean}[1]{\mbox{$\langle#1\rangle$}}
\newcommand{\scint}[2]{\mbox{$#1\!\times\!10^{#2}$}}

\begin{document}

\draft		

\title{Fluctuations in viscous fingering}

\author{Mitchell G.\ Moore\cite{mgmaddress},
Anne Juel\cite{ajaddress},
John M.\ Burgess,
W.\ D.\ McCormick, and 
Harry L.\ Swinney\cite{hlsaddress}}

\address{Center for Nonlinear Dynamics and Department of Physics\\
The University of Texas at Austin, Austin, Texas, 78712}
\date{October 16, 2001}
\maketitle

\widetext
\begin{abstract}
\vspace{-0.5in}

Our experiments on viscous (Saffman-Taylor) fingering in Hele-Shaw
channels reveal finger width fluctuations that were not observed in
previous experiments, which had lower aspect ratios and higher capillary
numbers Ca. These fluctuations intermittently narrow the finger from its
expected width. The magnitude of these fluctuations is described by a
power law, $\mathrm{Ca}^{-0.64}$, which holds for all aspect ratios
studied up to the onset of tip instabilities. Further, for large aspect
ratios, the mean finger width exhibits a maximum as Ca is decreased
instead of the predicted monotonic increase.
\end{abstract}
\begin{center}
\end{center}


\narrowtext


\vspace{-0.3in}

When a less viscous fluid displaces a more viscous fluid in a
Hele-Shaw channel (a quasi-2D geometry in which the width $w$ is much
greater than the channel thickness $b$), the interface between the
fluids forms a pattern of growing ``fingers''.  A single finger forms
at low flow rates; more complex branched patterns evolve at high flow
rates. This phenomenon is the prototypical example of moving interface
problems, such as dendritic growth and flame propagation, and thus
continues to receive attention for the insight it provides into these
important problems~\cite{Reviews}.  The problem was studied first by
Saffman and Taylor (1958)~\cite{Saffman/Taylor:1958}, who injected air
into oil in a Hele-Shaw cell and observed the formation of a single,
steadily moving finger whose width decreased monotonically to $1/2$ of
the channel width as the finger speed was increased. In subsequent
experimental~\cite{Tabeling/Zocchi/etc:1987},
numerical~\cite{DeGregoria/Schwartz:1986}, and
theoretical~\cite{McLean/Saffman:1981} work, the ratio of finger width
to channel width, $\lambda$, was found to depend on a modified
capillary number, $1/B=12\,(w/b)^2\,\mathrm{Ca}$, which combines the
aspect ratio, $w/b$, and the capillary number, Ca~$=\mu V/\sigma$,
where $\mu$ is the dynamic viscosity of the liquid, $V$ is the
velocity of the tip of the finger, and $\sigma$ is the surface
tension.  For large $1/B$ values, a transition to complex
patterns of tip-splitting occurs%
~\cite{Tabeling/Zocchi/etc:1987,HomsyGroup,Maxworthy:1987}.

Our experiments reveal fluctuations in the width of the evolving
viscous fingers that have not been reported in previous experiments%
~\cite{Saffman/Taylor:1958,Tabeling/Zocchi/etc:1987,HomsyGroup} nor
predicted theoretically.  The fluctuations intermittently
narrow the fingers from their expected width and are largest at low Ca,
falling off as a power law with increasing Ca. Further, at large
aspect ratios, the fluctuations are accompanied by substantial
departures from the monotonic dependence of the finger width on $1/B$
found previously%
~\cite{Tabeling/Zocchi/etc:1987,DeGregoria/Schwartz:1986,McLean/Saffman:1981}:
for aspect ratios $w/b \gtrsim 250$, which were not examined in
previous work, we find that the mean finger width no longer scales
monotonically with $1/B$; for smaller aspect ratios, our finger width
measurements are in accord with previous results.  Because large
aspect ratios should more closely approach the ideal of two-dimensional
flow, our observations pose a challenge to the assumptions underlying
theoretical analyses of viscous fingering.\newpage


\vspace*{0.94in}
{\it Experimental Methods.} We conducted experiments in a 254~cm long
channel formed of 1.9~cm thick glass plates. The
spacers between the glass plates were stainless steel strips with
thicknesses $b=0.051$~cm, 0.064~cm, 0.102~cm or 0.127~cm; the channel
width $w$ between the spacers was varied between 19.9~cm and
25.1~cm. Both glass plates were supported at the sides so that they
sagged similarly under their own weight (though by less than 0.1\% of
the gap depth). For aspect ratios under 150, we used a smaller channel
of length 102~cm and width 7.4~cm. Interferometric measurements
revealed that the root-mean-square variations in gap thickness were
typically 0.6\% or less in the large channel and 0.8\% in the small
channel.  Mechanical measurements of the bending of the glass due to
the imposed pressure gradient revealed that such deflections were
typically 0.2\% or less; the maximum deflection (2.2\%) was measured
in the widest channel close to the oil reservoir at the highest flow
rates.  Experiments were conducted with air penetrating a Dow Corning
silicone oil whose surface tension and dynamic viscosity was either
$\sigma=19.6$~dyne/cm, $\mu=9.21$~cP or $\sigma=20.6$~dyne/cm,
$\mu=50.8$~cP at laboratory temperature ($22^\circ$C). The oils wet
the glass completely.  A uniform flow rate was imposed by withdrawing
oil with a syringe pump from a reservoir at one end of the channel; an
air reservoir at atmospheric pressure was attached to the other end.


The channel was illuminated from below and images were obtained using a
camera and rotating mirror that captured 11 overlapping frames to produce
concatenated images of up to $1200 \times 10,000$ pixels at a resolution
of 0.25 mm/pixel. The interfaces were then digitally traced, yielding
finger width values accurate to 0.1\% in the larger channel and 0.3\% in
the smaller channel. For each flow rate, up to four time sequences of
20-30 digital interfaces were recorded.  Finger widths determined in
consecutive sequences agreed within the measurement accuracy.  Mean width
values agreed within 0.5\% for data sets repeated after channel
disassembly, cleaning, and reassembly.


For each interface image with a single, well developed

\newpage
\begin{figure}[tb]
\centerline{\includegraphics[width=3.375truein]{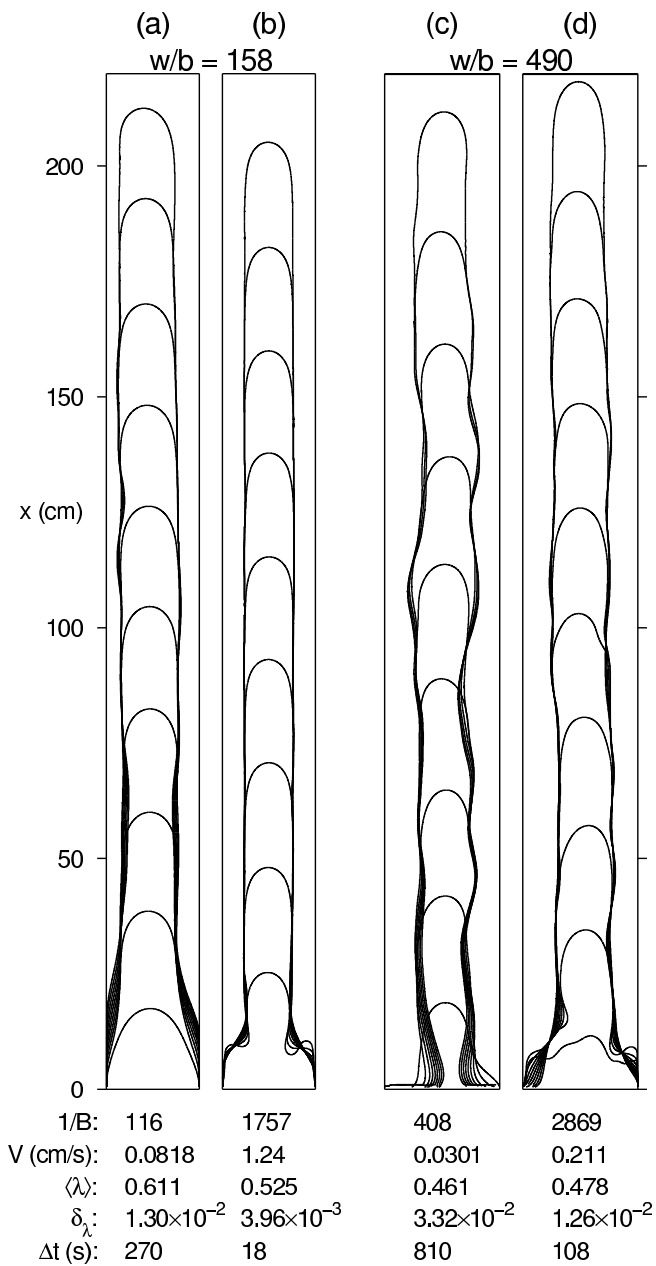}}
\vspace{0.5cm}
\caption{Finger images recorded at regular time intervals for different 
values of aspect ratio $w/b$ and modified capillary number $1/B$, with 
corresponding values of the tip velocity $V$, the mean finger width 
$\mean{\lambda}$, the rms fluctuation of the finger width 
$\delta_\lambda$, and the time $\Delta t$ between successive curves.}
\label{fig:time_series}
\end{figure}

\vspace{-0.7ex}\noindent finger, we found
the instantaneous finger width by averaging measurements taken in a narrow
window that was 5\% of the channel width beginning 1.2 channel widths behind
the tip. This measurement window was chosen to be small compared to the
length scale of the fluctuations (approximately a third of a channel
width); the results were not sensitive to the window's exact location.
From each time series of instantaneous widths, we determined the time
average $\mean{\lambda}$ and the root-mean-square (rms) fluctuation from
the mean $\delta_\lambda$. Each data set was analyzed for
flow rates up to the point of tip splitting, beyond which the finger
width $\lambda$ was no longer well defined~\cite{TipInst}.



{\it Results.} Typical interface image sequences are shown for $w/b=158$
and 490 in Fig.~\ref{fig:time_series}. For both aspect ratios the finger
width $\lambda$ fluctuates visibly at low flow velocities
(Fig.~\ref{fig:time_series}~(a),(c)).  In the smaller aspect ratio system
the width appears to become steady as the finger velocity is increased
(Fig.~\ref{fig:time_series}(b)), appearing exactly like the classic
``half-width finger'' of Saffman and Taylor.  However, with sufficient
resolution, fluctuations can still be measured for all velocities up to the
onset of tip instabilities.  In the higher aspect ratio system the width
fluctuates visibly for all flow rates (Fig.~\ref{fig:time_series}(d)). The
onset of tip instabilities in both cases occurs at $1/B \approx 4000$,
similar to values seen in previous
experiments~\cite{Tabeling/Zocchi/etc:1987,HomsyGroup}.

\begin{figure}[tb]
\centerline{\includegraphics[width=3.375truein]{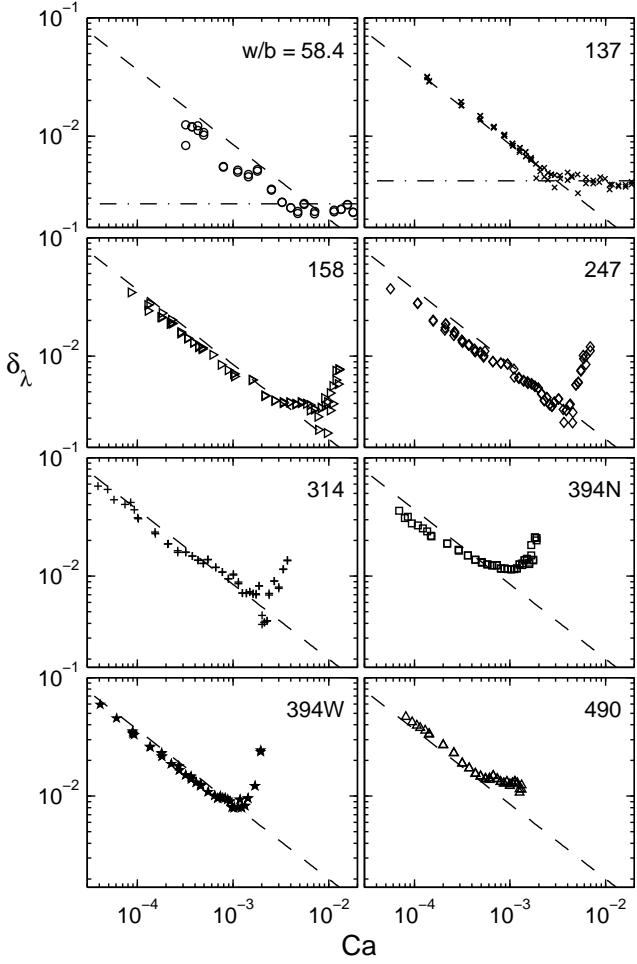}}
\vspace{0.5cm}
\caption{The rms fluctuation of finger width $\delta_{\lambda}$ as a 
function of capillary number Ca, where the dashed lines describe the best 
fit to all of the data sets within the observed scaling region, 
$\delta_{\lambda} = (\scint{1.1}{-4})\mathrm{Ca}^{-0.64}$.  (The data 
for 394N and 394W have the same aspect ratio but different widths; see 
caption of fig.~3.) The horizontal dash-dotted lines in the top two graphs 
correspond to the limits of measurement accuracy for that channel and 
geometry; this limit is below the lower edge of the graphs for the other 
data sets. Fluctuations with $\delta_\lambda \lesssim 10^{-2}$ are 
not obvious visually.  An upturn in $\delta_\lambda$ occurs at high Ca, 
signalling the onset of the secondary instabilities in the tips
of the fingers.}
\label{fig:dev_vs_Ca}
\end{figure}

For all the aspect ratios studied, we find that the rms fluctuation of
the finger widths is described by $\delta_\lambda =
A(\mathrm{Ca})$${}^{\beta}$ with $A = (1.1 \pm 0.3) \times 10^{-4}$ and
$\beta = -0.64 \pm 0.04$, as Fig.~\ref{fig:dev_vs_Ca} illustrates.
We also observe that
the instantaneous velocity of the finger tip fluctuates from the
average velocity; within the experimental uncertainty these velocity
fluctuations scale with Ca in the same manner as the width fluctuations.
Interestingly, a dependence of the form
$\mathrm{Ca}^{-2/3}$ appears frequently in theories of viscous fingering%
~\cite{Reviews,TL:86/PH:84}.

\begin{figure}[tb]
\centerline{\includegraphics[width=3.375truein]{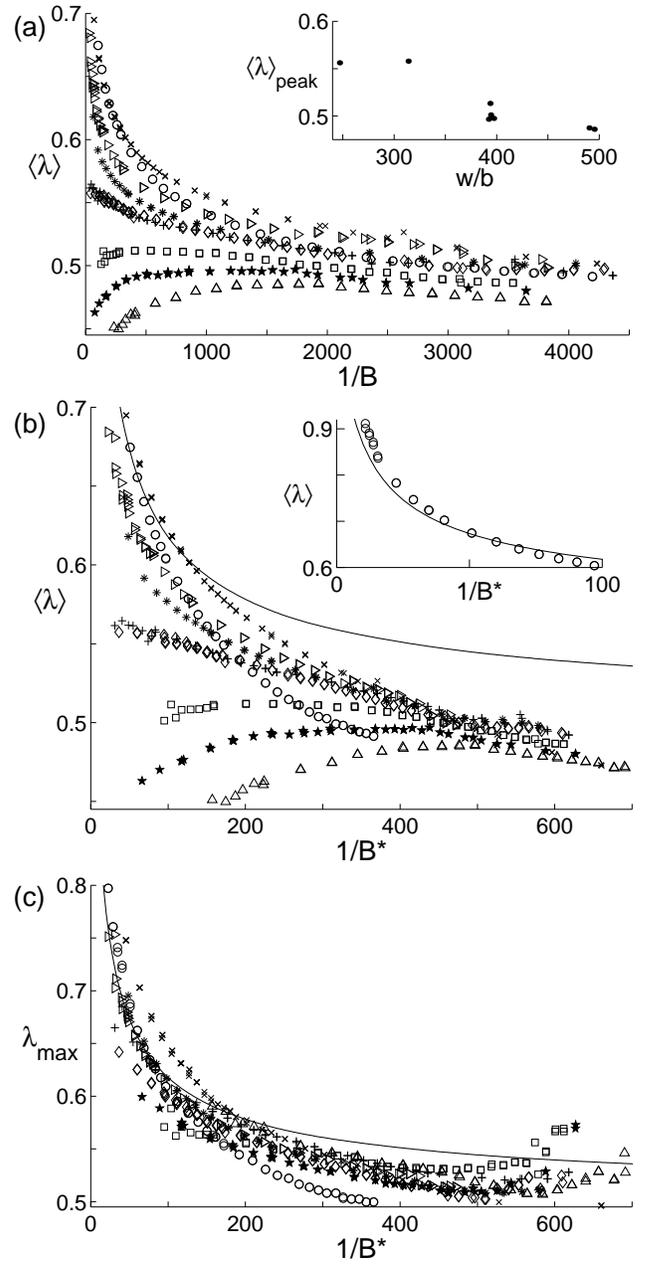}}
\vspace{0.5cm}
\caption{(a) Mean finger width $\mean{\lambda}$ {\it vs.\ }$1/B$ for
various aspect ratios $w/b$.  Higher aspect ratios give smaller
$\mean{\lambda}$ values.  The large $w/b$ data exhibit a peak width
$\mean{\lambda}_{\mathrm{peak}}$ as a function of $1/B$; the inset shows 
dependence of $\mean{\lambda}_{\mathrm{peak}}$ on aspect ratio $w/b$. (b) 
Mean finger width $\mean{\lambda}$ {\it vs.\ }$1/B^*$, which includes film 
wetting corrections.  The solid line is the theoretical curve of McLean 
and Saffman~[5].  The inset extends the $w/b=58.4$ curve to lower forcing. 
(c) Maximum width $\lambda_{\mathrm{max}}$ of the fluctuating finger 
width observed during evolution as a function of $1/B^*$.  Data for $w/b > 
137$ collapse roughly onto the same curve, which follows the 
McLean-Saffman theory more closely than the low aspect ratio results. 
Aspect ratio symbols:  $\circ$~58.4 ($w=$ 7.4~cm), $\times$~137 ($w=$ 
7.0~cm), $\triangleright$~158 ($w=$ 20.1~cm), $\ast$~214 ($w=$ 22.6~cm), 
$\Diamond$~247 ($w=$ 25.1~cm), $+$~314 ($w=$ 19.9~cm), $\Box$~394N ($w=$ 
20.0~cm), $\star$~394W ($w=$ 25.0~cm), $\bigtriangleup$~490 ($w=$ 
24.9~cm).}
\label{fig:width_vs_MCN}
\end{figure}

The fluctuations in finger width are accompanied by a substantial deviation
from the expected relation between finger width and velocity.  Our results
for the width of the viscous fingers for high $w/b$ are not described by
a single 
curve as a function of $1/B$ as predicted~\cite{McLean/Saffman:1981}, and 
the differences between data for different aspect ratios are far greater 
than those reported previously for low 
$w/b$~\cite{Tabeling/Zocchi/etc:1987}.  Figure~\ref{fig:width_vs_MCN}(a) 
illustrates this, showing the dependence of the mean finger width 
$\mean{\lambda}$ on $1/B$ for values of $w/b$ between 58.4 and 490.

In particular, the mean finger width exhibits a surprising maximum as the 
tip velocity is decreased for large aspect ratios
(Fig.~\ref{fig:width_vs_MCN}(a,b)). The value of $\mean{\lambda}$ at the 
peak, $\mean{\lambda}_{\mathrm{peak}}$, is plotted {\it vs.\ }$w/b$ in the 
inset of Fig.~\ref{fig:width_vs_MCN}(a).


To further compare our work with past results, we must account for the 
effects of the film of oil left behind on the plates as the finger 
advances.  To do so, we use a correction to the modified capillary number
introduced by Tabeling and Libchaber, $1/B^* = (1/B)/ \left[\pi/4 + 1.7\lambda
(w/b)(\mathrm{Ca})^{2/3} \right]$~\cite{TL:86/PH:84}.  Using this
correction, they obtained agreement between their
experimental measurements of $\lambda$ and the theoretical predictions
of McLean and Saffman~\cite{McLean/Saffman:1981} for $1/B^* < 100$
($1/B < 250$); we obtain similar agreement for $w/b = 58.4$, as the
inset of Fig.~\ref{fig:width_vs_MCN}(b) illustrates.
Fig.~\ref{fig:width_vs_MCN}(b) shows the dependence of the mean finger
width $\mean{\lambda}$ on $1/B^*$ for all aspect ratios. At high
$1/B^*$, the film wetting correction also collapses the data to a
common curve for all $w/b > 58.4$, though a slight displacement
downward of the large $w/b$ data remains.

\newpage


While our $\mean{\lambda}$ results do not exhibit the classical
scaling with $1/B^*$, the maximum value of the fluctuating finger
width observed during finger evolution, $\lambda_{\mathrm{max}}$,
does, as shown in
Fig.~\ref{fig:width_vs_MCN}(c). ($\lambda_{\mathrm{max}}$ is the
largest width value observed over time in a region behind the tip;
observing only near the tip excludes widening due to relaxation
effects.)  $\lambda_{\mathrm{max}}$ does not exhibit a peak with
decreasing $1/B$, and while $\lambda_{\mathrm{max}}$ has more
statistical noise than $\mean{\lambda}$, the $\lambda_{\mathrm{max}}$
data collapse onto similar, monotonically decreasing curves that agree
with the McLean-Saffman prediction at low $1/B^*$. The
$\lambda_{\mathrm{max}}$ data fall below the McLean-Saffman curve at
high $1/B^*$, but not as much as the data for $\mean{\lambda}$.

\newpage
\vspace{1in}

\noindent These
$\lambda_{\mathrm{max}}$ data suggest that the fluctuations represent
an intermittent narrowing of the fingers from their ``ideal'' width.


The finger width fluctuations and the peak in $\mean{\lambda}$ versus
$1/B$ have proven robust under variations of experimental conditions.
Both high and low viscosity oils gave the same results for the same
geometric configuration.  By treating the stainless steel spacers with
an anti-wetting agent, we changed the contact angle where the
interface is pinned at the back of the channel; this difference can be
seen between Fig.~\ref{fig:time_series}(b) and(c); we again found the
same results for the same geometric configuration.  To ensure that the
syringe pump was not imposing the observed velocity fluctuations, we
pumped one data set by gravity siphoning, again with identical results.
The effect of variations in the gap between the plates was examined for
several aspect ratios by over-clamping the channel along the sides, which
increases the intrinsic gap variations by a factor of 2.5 (measured
interferometrically); both the finger width fluctuations and the
location of $\mean{\lambda}_{\mathrm{peak}}$ with respect to $1/B^*$
were unchanged, although $\mean{\lambda}$ values decreased by about
4\% near the peak.

Though the fluctuation power law we observed remained unchanged for
{\it all} experimental variations, we did discover that measurements
on two channels with $w/b=394$ but different values of $w$ and $b$
yielded different results. The $\mean{\lambda}$ values for both
set-ups departed the common curve at approximately the same point with
decreasing $1/B^*$, but at lower $1/B^*$ the behavior was different --
compare 394W (394-wide) with 394N (394-narrow) in
Fig.~\ref{fig:width_vs_MCN}(b). This difference suggests that either
the width for a given $w/b$ and Ca is not unique, or perhaps a third
parameter, as yet unknown, is necessary to describe the problem for
high $w/b$ and/or low Ca.  Consistent with this are the
$\mean{\lambda}$ data for $w/b = 247$ and 314, which superpose closely
even though they have no geometric parameters in common.


It is unlikely that the fluctuations are caused by an instability of the
film wetting layer because the film is very thin at low capillary
numbers, where the
fluctuations are largest. Film wetting fluctuations would also cause
deviations in the growth rate of the area of the fingers, which we do
not observe.
We speculate that the fluctuations in finger width may be a
consequence of long-time relaxations of the interface at the back of
the channel, observable particularly in Fig.~\ref{fig:time_series}(c).
We also speculate that the peak in $\mean{\lambda}$
observed for decreasing $1/B$ at large aspect ratios may also occur at
small aspect ratios, but at values of $1/B$ too small to be reached.


In conclusion, we have discovered fluctuations that intermittently narrow
evolving single fingers; the magnitude of these fluctuations follows a
power law with the capillary number for all aspect ratios studied.  We
also have found a departure from the classic scaling of finger width versus
$1/B$ for large aspect ratios ($w/b \gtrsim 250$); the average finger
width narrows at low $1/B$, while the maximum finger width increases.
These phenomena are not predicted by existing viscous fingering
theories even though the phenomena are most pronounced for parameters
that more closely match the theoretical assumptions.


We thank A.\ Shaji for preliminary investigations, M.\ P.\ Brenner, M.\
Mineev and S.\ Tanveer for useful discussions, and J.\ B.\ Swift for
frequent guidance and advice.  This work was funded by the Office of Naval
Research, the NASA Microgravity Program, and the State of Texas
Advanced Technology Program (G.\ Carey, P.\ I.).


\end{document}